\pdfoutput=1
\documentclass[10pt,conference,letterpaper]{IEEEtran}
\usepackage{times,amsmath,epsfig}
\usepackage{graphicx}
\usepackage{amssymb}

\newcommand{\rar}{\rightarrow}

\newcommand{\mbb}{\mathbb}

\newcommand{\mca}{\mathcal}

\newcommand{\join}{\bowtie_\circ}
\newcommand{\disjoin}{\times_\circ}
\newtheorem{definition}{Definition}

\title{A Path Algebra for Multi-Relational Graphs}
\author{%
{Marko A. Rodriguez{\small $^1$}, Peter Neubauer{\small $^2$} }%
\vspace{1.6mm}\\
\fontsize{10}{10}\selectfont\itshape
$~^{1}$Graph System Architect, AT\&T Interactive\\
Santa Fe, NM 87506 USA\\
\fontsize{9}{9}\selectfont\ttfamily\upshape
marko@markorodriguez.com
\vspace{1.2mm}\\
\fontsize{10}{10}\selectfont\rmfamily\itshape
$~^{2}$VP Product Development, Neo Technology\\
21119 Malm\"{o}, Sweden\\
\fontsize{9}{9}\selectfont\ttfamily\upshape
peter.neubauer@neotechnology.com
}
\begin{document}
\maketitle
\begin{abstract} 
A multi-relational graph maintains two or more relations over a vertex set. This article defines an algebra for traversing such graphs that is based on an $n$-ary relational algebra, a concatenative single-relational path algebra, and a tensor-based multi-relational algebra. The presented algebra provides a monoid, automata, and formal language theoretic foundation for the construction of a multi-relational graph traversal engine.
\end{abstract}

\section{Introduction}

The adjacency of vertex $i$ and vertex $j$ is defined by the edge $(i,j)$. A structure of this form is called a graph and is usually defined as $\ddot{G} = (\ddot{V},\ddot{E})$, where $i,j \in \ddot{V}$ are vertices and $(i,j) \in \ddot{E}$ is the edge adjoining those vertices.\footnote{The ``high dot" notation denotes that $\ddot{G} \neq \dot{G} \neq G$, where $G$ is the main definition used throughout the article.} When the only distinguishing characteristic between two edges is the vertices they join, the graph is called single-relational. The reason for this is that there is only a single type of relation in the graph---namely, the binary relation $\ddot{E} \subseteq (\ddot{V} \times \ddot{V})$. Single-relational graphs have been used widely to model various systems of homogenous elements related by a single type of relation and as such, have numerous algorithms associated with their analysis \cite{netanal:brandes2005}. 

When the domain of discourse is variegated by a heterogeneous set of relations, then the multi-relational graph becomes the more applicable construct. A multi-relational graph can be defined as $\dot{G} = (\dot{V}, \dot{\mbb{E}})$, where $\dot{\mbb{E}}$ is a family of edge sets and $\dot{\mbb{E}}= \{\dot{E_1}, \dot{E_2}, \ldots, \dot{E_m} \subseteq (\dot{V} \times \dot{V})\}$. When $m > 1$, then there are multiple relations between the vertices of $\dot{V}$. Multi-relational graphs not only specify which vertices are adjacent to one another, they also specify the way in which they are adjacent. With respect to the formalisms of this article and without loss of generality, a multi-relational graph can also be represented as $G = (V, E)$, where $E$ is ternary relation, $E \subseteq (V \times \Omega \times V)$, and $\Omega$ is a set of edge labels (i.e.~relation types). Thus, in reference to the structure $\dot{G} = (\dot{V}, \dot{\mbb{E}})$, $|\dot{\mbb{E}}| = |\Omega|$ and $\sum_{n=1}^{n \leq |\dot{\mbb{E}}|} |\dot{E}_n| = |E| : \dot{E}_n \in \dot{\mbb{E}}$. The ternary relation model is the multi-relational graph structure used throughput this article. The reason for the use of this particular $G$ definition will be explained in \S \ref{sec:operations}.

Given the growing use of multi-relational graphs in computing \cite{dotslines:rodriguez2010} and the lack of graph techniques for such structures (relative to single-relational graphs), an algebraic model for traversing multi-relational graphs is presented. This article can be interpreted as a convergence of the $n$-ary relational algebra of \cite{rdbms:codd1970}, the concatenative single-relational path algebra in \cite{graphalg:russling1995}, and the multi-relational tensor algebra presented in \cite{pathalg:rodriguez2009}. However, unlike \cite{rdbms:codd1970}, the presented algebra is tied specifically to path construction by means of graph traversals as in \cite{pathalg:rodriguez2009} and \cite{graphalg:russling1995}. Next, unlike the algebra in \cite{graphalg:russling1995}, which is oriented primarily towards single-relational graphs, the presented algebra conveniently supports multiple relations as in \cite{rdbms:codd1970} and \cite{pathalg:rodriguez2009}. Finally, unlike \cite{pathalg:rodriguez2009}, the presented algebra is a concatenative, order-preserving variation of the relational algebra in \cite{rdbms:codd1970} and, as such, more aligned with \cite{graphalg:russling1995}.

The operations presented are summarized in the itemization below and are provided here as a consolidated summary for ease of reference.
\begin{itemize}
	\item $\|a\|$: the path length of path $a$.
	\item $\circ: E^* \times E^* \rar E^*$ : the concatenation of two paths.\footnote{The unary Kleene star operation $^*$ forms the free monoid $E^* = \bigcup_{n=0}^\infty E^i$, where $E^0 = \{\epsilon\}$ and $\epsilon$ is the empty/identity element.}
	\item $\sigma: E^* \times \mbb{N}^+ \rar E$: the projection of the $n^\text{th}$ edge of a path.
	\item $\gamma^-: E^* \rar V$: the projection of the tail (first element) of a path.
	\item $\gamma^+: E^* \rar V$: the projection of the head (last element) of a path.
	\item $\omega : E \rar \Omega$: the projection of the label of an edge.
	\item $\cup: \mca{P}(E^*) \times \mca{P}(E^*) \rar \mca{P}(E^*)$: the union of two path sets.
	\item $\join: \mca{P}(E^*) \times \mca{P}(E^*) \rar \mca{P}(E^*)$: the concatenative join of two path sets.
	\item $\disjoin : \mca{P}(E^*) \times \mca{P}(E^*) \rar \mca{P}(E^*)$: the concatenative product of two path sets.
\end{itemize}

Definitions of these operations are provided in \S \ref{sec:operations}. The use of these operations to represent basic traversal idioms is presented in \S \ref{sec:basic}. In \S \ref{sec:derivative}, regular paths can be recognized and generated as demonstrated in \S \ref{sec:recognizer} and \S \ref{sec:generator}, respectively. Making use of the algebra to evaluate single-relational graph algorithms is presented in \S \ref{sec:rich}. The algebra provides a set of core operations for constructing a multi-relational graph traversal engine that is founded on monoid, automata, and formal language theory.

\section{Core Operations\label{sec:operations}}

Traversing a graph is the process of moving over the edges specified in $E$. During a traversal, paths are derived and properties of those paths can be extracted.
\begin{definition}[Path]
A path $a$ in a multi-relational graph is a sequence, or string, where $a \in E^*$ and $E \subseteq (V \times \Omega \times V)$. A path allows for repeated edges. The path length is denoted $\|a\|$ and is equal to the number of edges in $a$. Any edge in $E$ is a path with a path length of $1$ as $e \in E \subset E^*$.
\end{definition}

The binary operation $\circ: E^* \times E^* \rar E^*$ is the concatenation of two paths into a new path such that if $(i,\alpha,j)$ and $(j,\beta,k)$ are two edges in $E$, then their concatenation is the path $(i,\alpha,j,j,\beta,k)$, where $i,j,k \in V$ and $\alpha,\beta \in \Omega$. Concatenation is associative (i.e.~$(a \circ b) \circ c = a \circ (b \circ c)$), not commutative (i.e.~it is generally true that $a \circ b \neq b \circ a$), and $\epsilon$ serves as an identity (i.e.~$\epsilon \circ a = a = a \circ \epsilon$).

Operations exist to extract information out of a path. The operation $\sigma: E^* \times \mbb{N}^+ \rar E$ is a projection that maps a path to the $n^\text{th}$ edge in that path. For example, if $a = (i,\alpha,j,j,\beta,k)$, then $\sigma(a,1) = (i,\alpha,j)$ and $\sigma(a,2) = (j,\beta,k)$. Next, for any path, $\gamma^-: E^* \rar V$ projects the tail (first vertex) of the path such that $\gamma^-((i,\alpha,j)) = i$. Likewise, $\gamma^+: E^* \rar V$, where $\gamma^+((i,\alpha,j)) = j$. Similarly, for edge labels, $\omega: E \rar \Omega$, where $\omega((i,\alpha,j)) = \alpha$.\footnote{All projection operations can be reduced to a single string indexing operation, but for the sake of clarity in the following discussion, they are presented as being atomic.}
\begin{definition}[Path Label]
The \textit{path label} of path $a$ is defined as the edge labels contained in $a$. Formally, if $a$ is a path, then the path label is constructed by $\omega': E^* \rar \Omega^*$, where, using concatenation,
\begin{equation*}
\omega'(a) = \prod_{n=1}^{n \leq \|a\|} \omega\left(\sigma\left(a,n\right)\right).
\end{equation*}
The path label of any single edge $e \in E$ is simply the edge's label as $\|e\| = 1$ and $\omega'(e) = \omega(\sigma(e,1)) = \omega(e)$.
\end{definition}

The binary operation $\cup: \mca{P}(E^*) \times \mca{P}(E^*) \rar \mca{P}(E^*)$ is standard set union. The binary operation $\join: \mca{P}(E^*) \times \mca{P}(E^*) \rar \mca{P}(E^*)$ is the concatenative join of two sets of paths such that if $A,B \in \mca{P}(E^*)$, then
\begin{align*}
A  \join B  = & \; \{ a \circ b \; | \; a \in A \; \wedge \; b \in B \\
		   & \;\; \wedge \; \left(a = \epsilon \; \vee \; b = \epsilon \; \vee \; \gamma^+(a) = \gamma^-(b) \right)\},
\end{align*}
where $\gamma^+(a) = \gamma^-(b)$ ensures that only \textit{joint} (i.e.~adjacent) paths are concatenated.\footnote{The defined concatenative join is analogous to the $\theta$-join in \cite{rdbms:codd1970}, where $\begin{array}{c}A \bowtie B \\ \gamma^+(a) = \gamma^-(b)\end{array}$. In this form, its known as an equijoin. A discussion relating concatenative join and the relational algebra is found in \cite{graph:pucheral1989}.} For example, if
\begin{equation*}
A = \left\{(i,\alpha,j),(j,\beta,k,k,\alpha,j) \right\}
\end{equation*}
and
\begin{equation*}
B = \left\{(j,\beta,j),(j,\beta,i,i,\alpha,k),(i,\beta,k)\right\},
\end{equation*}
then
\begin{align*}
A \join B =& \;  \{(i,\alpha,j,j,\beta,j),(i,\alpha,j,j,\beta,i,i,\alpha,k), \\
			 & \;\; (j,\beta,k,k,\alpha,j,j,\beta,j), \\
			 & \;\; (j,\beta,k,k,\alpha,j,j,\beta,i,i,\alpha,k) \},
\end{align*}
where $i,j,k \in V$, $\alpha,\beta \in \Omega$, and $(i,\alpha,j), (j,\beta,k), (k,\alpha,j), \\ (j,\beta,j), (j,\beta,i), (i,\alpha,k), (i,\beta,k) \in E$. Given that $\join$ is based on $\circ$, $\join$ is associative, but not commutative.
\begin{definition}[Path Jointness]
A path is joint is it satisfies the characteristic function $f: E^* \rar \{\top, \bot\}$ with the function rule
\begin{align*}
f(a) = 
	\begin{cases}
	  \top & \text{if } \|a\| = 1, \\
	  \top & \text{if } \bigwedge_{n=1}^{n<\|a\|-1} \gamma^+(\sigma(a,n)) = \gamma^-(\sigma(a,n+1)), \\
	  \bot & \text{otherwise}.
	\end{cases}
\end{align*}
The function maps to $\top$ if the path is joint and $\bot$ if it is disjoint.
\end{definition}
The binary operation $\join$ constructs joint paths. It may be the case that traversing disjoint paths is desirable.\footnote{For example, priors-based algorithms require the concept of ``teleportation" in order to make a disjoint jump in the graph.} The Cartesian product supports the concatenation of potentially disjoint paths. As such, $\disjoin: \mca{P}(E^*) \times \mca{P}(E^*) \rar \mca{P}(E^*)$, where $A \disjoin B = \{a \circ b \; | \; a \in A \; \wedge \; b \in B\}$.

Finally, to conclude this section, the reason why the $\dot{G}= (\dot{V}, \dot{\mbb{E}} = \{\dot{E_1}, \dot{E_2}, \ldots, \dot{E_m} \subseteq (\dot{V} \times \dot{V})\})$ definition of a multi-relational graph is not used is because when evaluating concatenative joins over binary relations, the edge label information is lost and thus, the path label can not be determined. In other words, if $e$ and $f$ are edges from two different binary relations, then $e \circ f$ would only provide a sequence of vertices and as such would not specify from which relations the join was constructed. This is a deficiency of the algebra in \cite{graphalg:russling1995}, where binary relations are used and $\circ: V^* \times V^* \rar V^*$ as opposed to $\circ: E^* \times E^* \rar E^*$, where $E = (V \times \Omega \times V)$. While the algebra in \cite{graphalg:russling1995} is applicable to multi-relational graphs (as any two relations can be joined), it was specifically intended for single-relational graphs, where problems involving path labels are not considered. In contrast, the specification defined in this article preserves path labels.

\section{Basic Traversals\label{sec:basic}}

From the explicit adjacencies (edges) defined in the edge set $E$, there exists implicit adjacencies (paths) defined by $e \circ f$, where $e,f \in E$ and $e \circ f \in E^*$. Given the previously defined operations, different types of common traversal idioms can be affected. 

\subsection{Complete Traversal}

All joint paths through a graph of length $n$ can be constructed using $\underbrace{E \join \ldots \join E}_{n \text{ times}}$. This type of traversal is called a complete traversal because there is no discrimination when joining except that the join vertex (i.e.~the head of the first path and tail of the second) be equal. When it is desirable to limit the set of paths derived by the traversal then the sets $A,B \subseteq E$ need to be defined and joined.

\subsection{Source Traversal}

A source traversal emanates from a particular set of vertices. Such a traversal is left restricting as it constructs paths whose tail vertex is an element of $V_s \subseteq V$. The first concatenative join must, on its left side, contain the set of all edges in $E$ that have their tail vertex in $V_s$. Therefore, when 
\begin{equation*}
A = \{ e \; | \; e \in E \; \wedge \; \gamma^-(e) \in V_s \},
\end{equation*}
$\underbrace{A \join E \ldots \join E}_{n \text{ times}}$ yields all joint paths of length $n$ emanating from the vertices in $V_s$. When $V_s = V$, a complete traversal is evaluated since $A = E$. For ease of expression, the complement of the set $V_s$ can be used to denote where \textit{not} to start a traversal from. For example, $\overline{V_s} = V \setminus V_s$ states to start the traversal from all vertices in $V$ except those in $V_s$.

\subsection{Destination Traversal}

A destination traversal is similar to a source traversal, except that it is right restricting as it constructs all paths of length $n$ whose head, or terminal, vertex is in $V_d \subseteq V$. In this way, when
\begin{equation*}
B = \{ e \; | \; e \in E \; \wedge \; \gamma^+(e) \in V_d \},
\end{equation*}
$\underbrace{E \join \ldots E \join B}_{n \text{ times}}$ is a destination traversal. When $V_d = V$, a complete traversal is evaluated because $B = E$ in such situations.

By combining a source and destination traversal, its possible to emanate from particular vertices and arrive at particular vertices, where $\underbrace{A \join E \ldots E \join B}_{n \text{ times}}$ is the set of all joint paths that start from vertices in $V_s$, end at vertices in $V_d$, and are of length $n$. Source and destination traversals can also be used to ensure that each edge in the path goes through a particular set of vertices by specifying, at some particular $\join$ step, the source (or destination) vertex set as $V_s$ (or $V_d$) before enacting the next concatenative join.

\subsection{Labeled Traversal}

A traversal can be constrained to particular path labels by defining an edge set that is a function of its edge labels. For example, if $\Omega_e \subseteq \Omega$, $\Omega_f \subseteq \Omega$,
\begin{equation*}
A = \{ e \; | \; e \in E \; \wedge \; \omega(e) \in \Omega_e \},
\end{equation*}
and
\begin{equation*}
B = \{ f \; | \; f \in E \; \wedge \; \omega(f) \in \Omega_f \},
\end{equation*}
then $A \join B$ denotes all paths where $\omega(\sigma(a,1)) \in \Omega_e$ and $\omega(\sigma(a,2)) \in \Omega_f$. When $\Omega_e = \Omega_f = \Omega$, a complete traversal is enacted as, in such situations, $A = B = E$. The labeled traversal is possible because the relation type is represented in the edge definition $E \subseteq (V \times \Omega \times V)$ and there exists the label projection function $\omega: E \rar \Omega$. 

\section{Derivative Traversals\label{sec:derivative}}

The basic traversals defined in \S \ref{sec:basic} can be mixed and matched to yield different types of joint paths in $E^*$. This section will introduce some typical applications of the presented multi-relational path algebra to problems that are specific to multi-relational graphs---focusing primarily on problems involving regular paths.\footnote{For the sake of simplicity, only regular paths are discussed. However, with more machinery (e.g.~memory structures), more complex traversals can be expressed using the core operations presented in \S \ref{sec:operations}.}

\subsection{Regular Path Recognizer\label{sec:recognizer}}

The presented multi-relational path algebra has application to regular expressions and their corresponding finite state automata. Before presenting this application, an example-specific set-builder notation is introduced in order to specify subsets of $E$ in a more concise, readable manner than previously presented. A source edge set can be specified as $[i,\_,\_] \equiv \bigcup_{\alpha \in \Omega} \bigcup_{j \in V} (i,\alpha,j) : (i,\alpha,j) \in E$ in order to denote the set of all edges that emanate from vertex $i$. A destination edge set can be specified as $[\_,\_,j] \equiv \bigcup_{i \in V} \bigcup_{\alpha \in \Omega} (i,\alpha,j) : (i,\alpha,j) \in E$ in order to denote the set of all edges that terminate at vertex $j$. A labeled edge set can be specified as $[\_,\alpha,\_] \equiv \bigcup_{i \in V} \bigcup_{j \in V} (i,\alpha,j) : (i,\alpha,j) \in E$ in order to denote the set of all edges that have $\alpha$ as their label. Finally $[\_,\_,\_] = E$.

\begin{figure}[h!]
	\centering
	\includegraphics[width=0.475\textwidth]{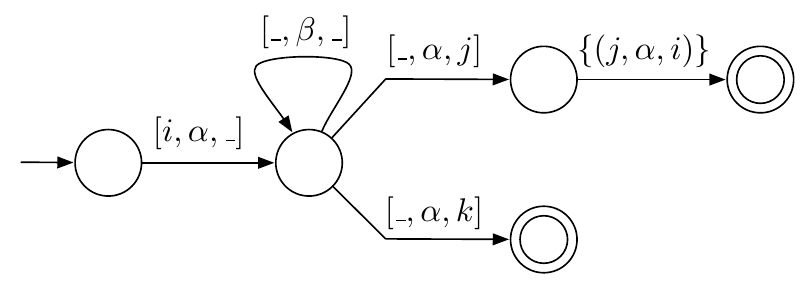}
	 \caption{\label{fig:fsm} A finite state automaton to recognize and generate a set of paths in $\mca{P}(E^*)$. The left most state is the start state and the double-circle states denote accepting states.}
\end{figure}

If $E$ is the regular expression alphabet, then $\emptyset$, $\epsilon$, and any $e \in E$ are regular expressions. If $R$ and $Q$ are regular expressions, then $R \cup Q$, $R \join Q$, and $R^*$ are regular expressions \cite{comput:moret1997}.\footnote{The $\disjoin$ operation can be used to recognize potentially disjoint paths, but in practice, when only joint paths are being recognized then $\join$ is a more efficient use of resources as $R \join Q \subseteq R \disjoin Q$.} A regular expression over $E$, and corresponding finite state automaton, recognize a set of joint paths in $\mca{P}(E^*)$.\footnote{The common operations $R^+$, $R?$, and $R^n$ used in practice can be represented as $R \join R^*$, $R \cup \{\epsilon\}$, and $\underbrace{R \join \ldots \join R}_{n \text{ times}}$, respectively.} For example,
\begin{equation*}
[i,\alpha,\_] \join [\_, \beta, \_]^* \left(\left([\_, \alpha, j] \join \{(j,\alpha,i)\}\right) \; \cup \; [\_, \alpha, k]\right)
\end{equation*}
recognizes all paths emanating from $i$, terminating at $i$ or $k$, with the first and last label traversed being $\alpha$, and all intermediate edge labels (zero or more) being $\beta$. The corresponding finite state automaton is diagrammed in Figure \ref{fig:fsm}, where the transition function is based on set membership, not equality.\footnote{Given that set membership can be represented element-wise as element equality under \textit{or}, each element of the transition label edge set can be individually denoted as a transition with the same tail and head state. As such, the typical finite state automaton transition exists. For diagram clarity, set membership is used instead of equality.} 

Regular paths in graphs are explored in depth in \cite{path:mendelzon1989}, where only paths with particular path labels are considered for recognition. In other words, in \cite{path:mendelzon1989}, a regular expression is defined for the alphabet $\Omega$, where above, its defined for $E$.

\subsection{Regular Path Generator\label{sec:generator}}

By making use of a non-deterministic single-stack automaton with a stack alphabet of $\mca{P}(E^*)$, it is possible to generate all paths in $G$ that can be recognized by some regular expression. The non-deterministic aspect of the automaton ensures that all branches in the state machine are taken ``in parallel." The single-stack aspect refers to the fact that the automaton (and thus, its cloned/branched automata) maintain a first-in/last-out stack memory that can be pushed and popped.

Initially, the automaton's stack contains the element $\{\epsilon\}$. The automaton will halt whenever its stack element is $\emptyset$ or is in an accepting state. For each state transition (which happens unless the automaton has been halted), the path set defined on the transition label is joined on the right with the path set popped off the stack. The result of the join is then pushed back onto the stack. Whenever a branch in the automaton's state graph is approached, all branches are taken ``in parallel." Thus, given the automaton diagrammed in Figure \ref{fig:fsm}, the following joins are evaluated.
\begin{align*}
& \{\epsilon\} \join [i,\alpha,\_] \join [\_, \alpha, j] \join \{(j,\alpha,i)\} \\
& \{\epsilon\} \join [i,\alpha,\_] \join [\_, \alpha, k] \\
& \{\epsilon\} \join [i,\alpha,\_] \join [\_,\beta,\_]  \ldots \join [\_, \alpha, j]  \join \{(j,\alpha,i)\} \\
& \{\epsilon\} \join [i,\alpha,\_] \join [\_,\beta,\_]  \ldots \join [\_, \alpha, k] \\
\end{align*}
The union of the first (and only) element of all the stacks across all branches of accept-state automaton forms the set of all paths in $G$ that satisfy the regular expression.

\subsection{Constructing Semantically-Rich Single-Relational Graphs\label{sec:rich}}

Most of the graph algorithms in existence today have been developed for single-relational graphs. Examples of such algorithms include the geodesics (e.g.~closeness centrality, betweenness centrality), spectral (e.g.~eigenvector centrality, spreading activation), and assortative (e.g.~scalar and discrete) algorithms (see \cite{netanal:brandes2005} for a consolidate review and analysis of many such algorithms). When applied to multi-relational graphs, these algorithms have the potential drawback of losing their meaning and thus, their applicability. To explicate this statement, it is important to consider the way in which a single-relational graph algorithm can be formally applied to multi-relational graphs. One method that can be employed is to simply ignore edge labels and, potentially, repeated edges between the same two vertices. However, when there are numerous ways in which one vertex can be related to another vertex, what is the resulting semantics of, say, a centrality algorithm? Another method is to extract a single edge relation, based on its label, from the multi-relational graph. For example, its possible to construct the binary edge set
\begin{equation*}
E_\alpha = \{ (\gamma^-(e),\gamma^+(e)) \; | \; e \in E \; \wedge \; \omega(e) = \alpha \}
\end{equation*}
and utilize that subgraph as the source of a single-relational graph algorithm. However, with multiple ways in which vertices can be related, more abstract relationships can be inferred through paths. Thus, in the final method, single-relational graphs can be generated from the multi-relational graph through the derivation of implicit edges defined through paths. Using a simple example, if $\alpha,\beta \in \Omega$ are two edge labels, then all $\alpha\beta$-paths can be constructed when $A = \{ e \; | \; e \in E \; \wedge \; \omega(e) = \alpha \}$, $B = \{ e \; | \; e \in E \; \wedge \; \omega(e) = \beta \}$ and $A \join B$. The tail and head vertices of these paths can then be projected to form a new binary edge set
\begin{equation*}
E_{\alpha\beta} = \bigcup_{a \in A \join B} \left(\gamma^-(a), \gamma^+(a)\right).
\end{equation*}
Thus, $E_{\alpha\beta} \subseteq (V \times V)$ can be subjected to all known single-relational graph algorithms. For regular paths, a regular path generator can be used as in \S \ref{sec:generator}. Mapping single-relational graph algorithms over to the multi-relational domain is explored in depth in \cite{pathalg:rodriguez2009}.

\section{Conclusion}

This article defined a path algebra for multi-relational graphs represented as $G = (V,E \subseteq (V \times \Omega \times V)$. The core traversal types (complete, source, destination, and labeled) allow for the expression of more expressive traversals through the restriction of the join set $E$. Applications to regular path recognizers (\S \ref{sec:recognizer}), generators (\S \ref{sec:generator}), and ``semantically-rich" single-relational graph construction (\S \ref{sec:rich}) were presented. Generally, the algebra has applicability to the construction of a multi-relational graph traversal engine.

\vfill


\end{document}